# Towards Understanding the Benefits and Challenges of Demand Responsive Public Transit - A Case Study in the City of Charlotte, NC

Sanaz Sadat Hosseini[a,*], Mona Azarbayjani[b], Jason Lawrence[c] and Hamed Tabkhi[d]

[a]Department of Civil and Environmental Engineering, The University of North Carolina at Charlotte, Charlotte, NC, USA
[b]School of Architecture, The University of North Carolina at Charlotte, Charlotte, NC, USA
[c]Charlotte Area Transit System (CATS), Charlotte, NC, USA
[d]Department of Electrical and Computer Engineering, The University of North Carolina at Charlotte, Charlotte, NC, USA

ABSTRACT

Access to reliable public transportation is essential for addressing socio-economic disparities, particularly in low-income communities that rely heavily on transit for accessing jobs, healthcare, and essential services. This study investigates the challenges faced by transit-dependent populations in Charlotte, NC, focusing on the spatial and service-related inequities within the current public bus system. Our research initially evaluates critical issues such as extended wait times, unreliable schedules, and limited accessibility, which significantly impact the daily lives of low-income residents. In response to these challenges, we gathered data to assess the potential for a connected, demand-responsive bus system designed to minimize transit gaps and enhance service efficiency in the future. This evaluation included an analysis of the existing Charlotte Area Transit System (CATS) mobile applications and the exploration of user acceptance for a proposed smart, on-demand transit technology. Through surveys conducted across key bus lines—including the Sprinter line and Bus Lines 7, 9, and 97-99—we identified significant shortcomings in the current system. However, our findings also indicate a strong willingness among participants to adopt new transit solutions, provided that they effectively address current issues and alleviate concerns related to smartphone accessibility, privacy, and trust. This research contributes valuable insights into the modernization of public transit systems in Charlotte, highlighting the importance of user-centric approaches in developing innovative, equitable, and efficient transportation solutions.

## 1. Introduction

The growing interest in Demand Responsive Transit (DRT) has sparked discussions on its potential benefits and challenges, particularly in low-income neighborhoods where traditional public transit often fails to meet residents' needs [1]. DRT systems, which utilize flexible routing and scheduling technology, promise to enhance accessibility, reduce wait times, and offer more personalized service options compared to fixed-route buses [2]. However, the effectiveness of DRT systems in addressing transit inequities, particularly in car-oriented cities, remains a subject of ongoing research. This study aims to explore these issues by focusing on two main topics: the willingness of different population groups to use various transport apps and the unique challenges faced by low-income neighborhoods in potentially adopting DRT solutions as part of the future of public transit systems.

Understanding the dynamics of public transit use and the potential of DRT systems requires a deep dive into specific contexts where these systems could be implemented. In low-income neighborhoods, where residents often rely heavily on public transportation for access to jobs, healthcare, and other essential services, the introduction of DRT could either alleviate or exacerbate existing transit disparities [3]. Therefore, evaluating the willingness to use transport apps and the impact of DRT on these communities is crucial for developing effective, equitable public transit solutions.

Figure 1 shows disparities between high ridership residential locations and transit service availability in four major cities in the Southeast, a fast-growing region characterized by auto-centric urban forms [4]. For underserved neighborhoods in Charlotte, Charleston, Nashville, and Jacksonville to meet benchmarks for adequate service in neighborhoods with similar characteristics, a reduction in average wait times of 50, 46, 219, and 40 minutes is required. A total of 33%, 39%, 38%, and 20% of these cities' populations are underserved ridership households. Additionally, COVID-19 has drawn attention to transit gaps between home and work, especially for essential workers [5].

Charlotte, North Carolina, serves as the case study for this research. The city is one of the fastest-growing in the United States, with a history of car-oriented, low-density suburban development that has contributed to significant public transportation challenges. Despite expansions in light rail service, Charlotte's bus system remains the primary mode of transportation for many low-income residents, who face long wait times and multiple transfers to reach job-dense areas. Charlotte also ranks fiftieth out of the fifty largest U.S. cities in terms of socio-economic mobility [6]. The inefficiencies of the current bus system—where headways on half the routes exceed 45 minutes—disproportionately affect low-income residents who rely on public transit to reach employment centers located across the city [7].

*Corresponding author

shossei7@charlotte.edu (S.S. Hosseini); mazarbay@charlotte.edu (M. Azarbayjani); Jason.Lawrence@charlottenc.gov (J. Lawrence); htabkhiv@charlotte.edu (H. Tabkhi)

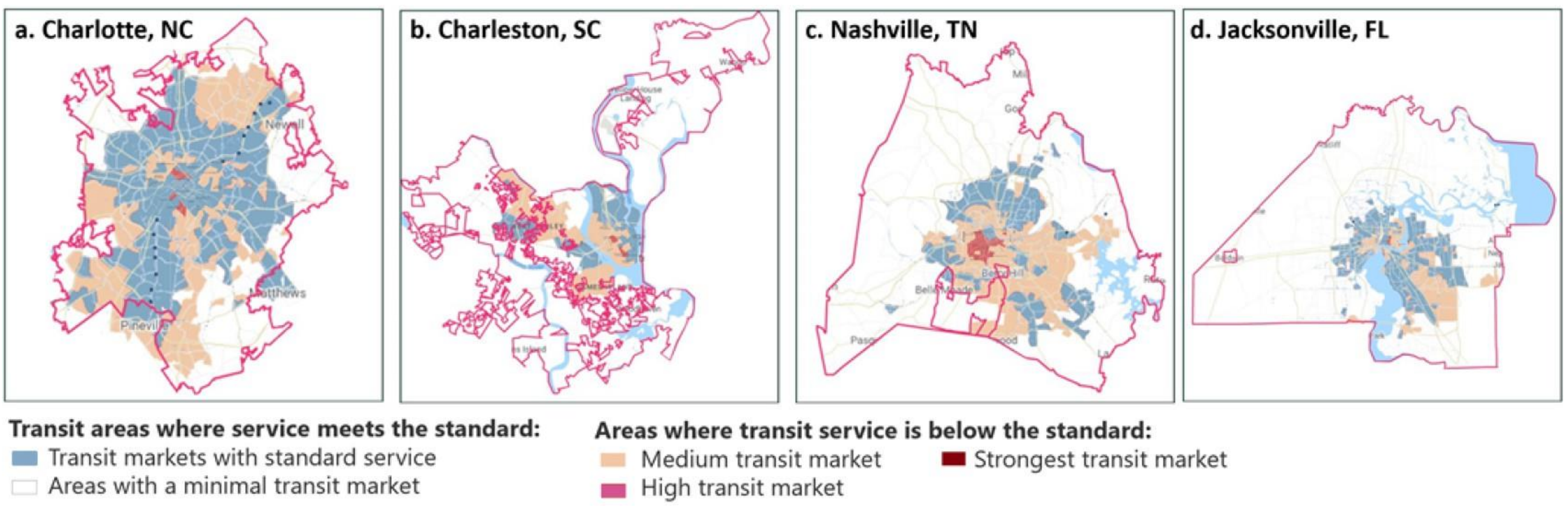

**Figure 1:** Mismatch between transit market and transit service available to the communities for four rapidly growing cities in southeast

It is evident that the lack of adequate public transportation is a common problem in growing cities like Charlotte, and addressing these issues is critical to preventing future social and economic crises. To tackle the spatial mismatch in Charlotte—where some areas lack critical services and adequate public bus service to job-dense areas—this study aims to understand the demand-supply gap in public transportation, particularly in low-income communities [7, 8]. This research also includes a user experience study, collecting feedback from bus passengers to assess the potential of on-demand technology and DRT as viable future solutions for Charlotte.

In many low-income housing neighborhoods, the lack of access to food, healthcare, and essential services further suppresses upward mobility and exacerbates existing disparities [9]. Early findings suggest that DRT systems could potentially enhance the quality of life for these communities and reduce disparities in these areas. Through in-person surveys, questionnaires, and town hall meetings with residents, neighborhood leaders, and stakeholders, this research have studied the potential impact of DRT in reducing bus ridership problems and improving access to essential services in Charlotte's low-income neighborhoods and the broader community.

To comprehensively explore the potential of DRT systems as a future solution for Charlotte's public transit, in this study we examined a diverse range of bus lines across the city. These routes have been carefully selected to represent various geographical contexts, including suburban and urban areas, as well as routes with differing levels of ridership—from low to high. We have focused on six bus routes in Charlotte, including the Airport Sprinter line, which connects City Center to the Charlotte Douglas International Airport, primarily serving essential workers from low-income areas. The study also examined bus lines 7 and 9, which run from Uptown Charlotte to North and East Charlotte, providing access to urban amenities along major thoroughfares. Additionally, the research included bus lines 97-99, known as the North Meck Village Rider, which connects Huntersville, Cornelius, and Davidson to key transit hubs and offers flexible routing for passengers who schedule trips in advance.

By conducting surveys and engaging with residents who rely on these bus lines, we aimed to gather valuable insights into their experiences, needs, and openness to new transit solutions like DRT. This approach allowed us to understand the varying perspectives of different communities, helping to identify the potential challenges and opportunities for implementing more equitable and efficient public transit options in the future. Ultimately, this study seeks to inform the conversation on how DRT could be tailored to meet the specific transportation needs of Charlotte's diverse population.

The remainder of this paper is structured as follows. Section 2 provides a detailed literature review, highlighting previous research on Demand Responsive Transit (DRT) and its implications for public transportation in other cities. Section 3 discusses the city of Charlotte as the case study of this research and current state of the Charlotte Area Transit System (CATS), focusing on its operations, technology integration, and challenges. Section 4 outlines the research methodology used in this study, including the selection of bus lines, survey design, and data collection methods. Section 5 presents the results of the survey, segmented into urban and suburban lines, and discusses key findings related to demographics, travel quality, and technology adoption. In Section 6, we delve into the discussion, exploring the implications of the findings for the future of DRT and public transit in Charlotte and beyond. Section 7 concludes the paper by summarizing the key contributions of this study and offering directions for future research and policy development.

## 2. Literature Review and Background

In this section, we examine the successful implementation of technology-driven approaches to enhance the efficiency of public transit systems, with a particular focus on smart, data-driven solutions and the role of user acceptance in the success of these technologies. As cities worldwide transition to smarter transportation systems, the demand for real-time information and responsive services is growing among citizens. Demand Responsive Transit (DRT) has emerged as a promising solution to bridge gaps in public transit by leveraging technology to provide more flexible, user-centric services. However, the effectiveness of these solutions is closely tied to the use and acceptance of transit apps by the public. Understanding how users interact with these digital platforms and their willingness to adopt such technologies is crucial for the widespread implementation of innovative transit solutions [10, 11, 12, 13, 14, 15].

While early research has predominantly focused on rail transit, recent studies have begun to explore demand-responsive public bus transit, leveraging data analytics to optimize scheduling and minimize passenger wait times [15, 16, 17, 18, 19, 20, 21, 22]. The integration of General Transit Feed Specification (GTFS), Automatic Vehicle Location (AVL), and Automatic Passenger Counter (APC) technologies facilitates data collection for analysis in these advanced systems [23, 24, 25, 26, 27]. However, these technologies are only available in a fraction of public transportation systems and are often not utilized in real-time to adjust demand and supply dynamically. Consequently, static scheduling and load balancing remain prevalent, offering limited insight into real-time transit demand distribution [23, 25].

The application of these technologies has been explored in various contexts. In Singapore, [21] explored Mobility-on- Demand ride-sharing services applied to high-capacity buses in densely populated areas. The study focused on Dynamic Bus Routing (DBR) and developed a simulator using a modified insertion algorithm to model the dynamic routing of buses. The project demonstrated that dynamically routed buses could be an efficient mode of mass transit, potentially offering significant advantages over existing fixed routes. This aligns with the objectives of our research.

Similarly, Nannapaneni et al. (2019), [22] investigated the rerouting of a single bus under varying travel demands. The study proposed a flexible framework for public transit rerouting to better serve spatially and temporally changing travel demands. This framework was demonstrated on Route 7 of the Nashville Metropolitan Transit Authority (MTA), showing that flexible routes could reduce additional travel time without exceeding the existing slack time in static schedules. This approach also considered the percent increase in travel demand to analyze rerouting effectiveness.

AI-controlled on-demand bus systems have also emerged as a promising solution to urban transportation challenges. In Japan, Next Mobility Co Ltd (Next Mobility JV) operates AI-controlled on-demand buses that generate routes in real-time based on passenger requests submitted through smartphone apps. These buses use deep learning to collect operational data on rider destinations and traffic conditions, enabling more efficient operations over time. This smart system allows passengers to book rides and pay using their smartphones, promoting a shift from private car use to public transit, thus fostering more sustainable urban transportation [10].

In the U.S., cities like Wilson, North Carolina, and Dallas, Texas, have replaced traditional fixed-schedule transit services with on- demand services, allowing residents to summon rides via an app. The case of DART's GoLink service in Dallas revealed that while on-demand services might be more expensive per rider than traditional buses, they are more cost-effective than operating routes with low ridership [28].

## 2.1. Transit Apps' Use and Acceptance

The success of smart public transit solutions, including DRT systems, is closely tied to the use and acceptance of transit apps by the public. Understanding how users interact with these digital platforms and their willingness to adopt such technologies is crucial for the widespread implementation of innovative transit solutions.

Harmony & Gayah (2017) [29] emphasize the importance of user-centered design in the development of transit apps, noting that user satisfaction is significantly influenced by the app's usability and the quality of real-time information it provides. They argue that for transit apps to be effective, they must be designed with a deep understanding of the users' needs and preferences. This user- centric approach ensures that the technology not only meets the functional requirements but also enhances the overall user experience, which is key to increasing adoption rates [29].

Mulley et al. (2017) [30] explore the role of digital platforms in transforming public transportation services. They highlight how these platforms can improve service delivery by offering personalized and real-time information, thereby increasing customer satisfaction. Their study also discusses the potential barriers to the adoption of these technologies, such as privacy concerns and the digital divide, which must be addressed to ensure that all user groups can benefit from the advancements in transit technology [30].

Romero et al. (2022) [31] investigate the factors influencing the adoption of mobile transit apps, particularly in the context of smart cities. Their findings indicate that perceived ease of use, perceived usefulness, and social influence are significant predictors of user acceptance. Additionally, the study highlights the importance of trust in technology, which can be fostered through transparent data management practices and reliable app performance. The authors suggest that addressing these factors can lead to higher adoption rates, making transit apps a vital component of modern public transportation systems [31].

The integration of these insights into the broader context of smart, demand-responsive public transit is essential for developing a holistic understanding of how to effectively implement such systems. As smart technologies continue to evolve, the role of transit apps in facilitating seamless and efficient transit experiences will only grow in importance.

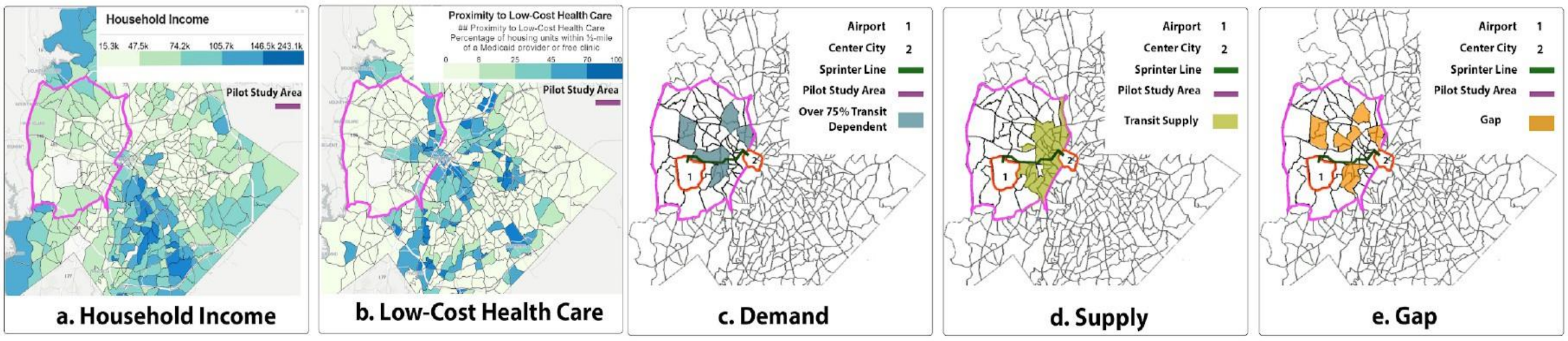

**Figure 2:** a. Household Income, b. Healthcare Proximity, c. Transit Dependent Population, d. Public Transport Supply, e. Supply-Demand Gap

Given the technological advancements and user considerations highlighted in this literature review, smart on-demand bus systems, real-time data, and optimized scheduling represent innovative approaches for creating efficient and user-friendly transportation systems. Similarly, this study examines six CATS bus routes in Charlotte, North Carolina, to understand how these technologies might help low-income communities and other riders utilize CATS buses more effectively in future. It is essential to consider that any technological innovation should be studied from two closely linked perspectives: the technological tools themselves and the users who must accept, adopt, and utilize these tools [32, 33, 34, 35]. To facilitate the implementation and adop- tion of smart technologies, it is crucial to examine the user's perspective, which will ultimately determine the success of these innovations [36].

Therefore, before advancing to the technological aspects of this research that will be provided in our future articles and publications, in this paper we plan to present and analyze the perspectives of community stakeholders and individuals regarding their needs, concerns, and reactions to the proposed technology. Understanding these perspectives is critical for the successful adoption and implementation of new transit solutions in Charlotte.

# 3. Case Study: Bus Transit and Technology Integration in Charlotte

## 3.1. Overview of Bus Transit in the U.S.

Public transportation systems across the United States vary in their offerings, tailored to meet the needs of different communities, streets, and neighborhoods. Depending on the street context and service needs, various design elements can complement these services [23]. Nearly every major U.S. city offers some form of bus service, many operating 24 hours a day, with flexible routes and frequent stops to provide accessible transit options for all areas within a community. Bus transit in the U.S. can be categorized into several route types, each serving different purposes and contexts within urban areas:

- **Downtown Local Routes:** Provide core transit functions for short distances within high-demand areas, such as downtowns. Often operate parallel to longer routes, with high stop frequencies of four or more per mile [23].
- **Local Routes:** Balance access and speed, typically with 3-5 stops per mile. Used for short- to medium-length trips, often within or between neighborhoods, downtowns, and other hubs [23].
- **Rapid Routes:** Designed for longer trips or high-demand corridors, rapid routes feature fewer stops (1-3 per mile) and can operate as trunk lines or on the same routes as local services but with limited stops [23].
- **Coverage Routes:** Serve low-density areas or regions with poorly connected street networks, often added as deviations to local routes to cover small ridership pockets. Stop frequencies range from 2 to 8 per mile [23].
- **Express Routes:** Provide direct point-to-point service, typically using limited-access highways with non-stop express segments. These routes are concentrated during peak periods, offering less frequent but faster service for longer distances [23].

## 3.2. Charlotte Area Transit System (CATS)

### *3.2.1. CATS Operations and Ridership*

The Charlotte Area Transit System (CATS) is the primary public transportation provider in the Charlotte metropolitan area, operating bus and rail services throughout Mecklenburg County and surrounding areas. Established in 1999, CATS carries approximately 320,000 riders on an average week, with bus routes having an average stop distance of about 0.2 to 0.25 miles in the urban core, extending to 0.5 miles or more in suburban areas [37, 38].

CATS operates 73 different bus routes throughout Mecklenburg County, including cities and towns such as Charlotte, Davidson, Huntersville, Cornelius, Matthews, Pineville, and Mint Hill. Weekly, CATS buses serve about 190,000 passengers [39, 40]. The types of routes operated by CATS include:

- **Local Routes:** Operate primarily within Charlotte and Mecklenburg County, often starting at the Charlotte

Transportation Center (CTC) in Uptown Charlotte and connecting to various neighborhoods. Bus line 7 (Beatties Ford) and bus line 9 (Central Avenue) are examples of local routes.

- **Express Routes:** Serve areas like Union County, northern Mecklenburg County, the Lake Norman area, Gastonia, Rock Hill, and parts of South Carolina.

- **Sprinter Bus Line:** This rapid transit line connects the Charlotte Transportation Center (CTC) to Charlotte Douglas International Airport, providing a direct link from the city center to the airport. The Sprinter line is expected to be replaced by the Lynx Silver Line upon its completion.

- **Low-Ridership Routes:** Examples include the North Meck Village Rider buses (lines 97-99), serving areas like Cornelius, McCoy Road, and Huntersville. CATS is exploring options to transform these routes into more efficient, on-demand services.

- **Special Transportation Service (STS):** A paratransit service offering transportation to individuals with disabilities, ensuring they have access to the same locations and times as the fixed-route bus services.

- **Future Plans:** CATS is proposing the I-77 Bus Rapid Transit service to connect northern Mecklenburg and southern Iredell counties to Uptown Charlotte [37].

Charlotte's radial bus system poses significant challenges for bus riders, particularly those from low-income neighborhoods. These residents often must travel first to the Center City, where job density is high but concentrated around financial and banking institutions, before transferring to other lines that serve low-skill, low-wage areas. This system disproportionately affects low- income communities, who rely heavily on public transportation for access to employment and essential services [10, 7].

As shown in Figure 2, part e, the west of Charlotte is particularly affected by significant transit disparities. This area is home to many low-income families who face poor proximity to healthcare and other amenities. These disparities, calculated as the difference between transit-dependent populations and public transit supply, are further exacerbated by a lack of access to affordable healthcare and other essential services. The data, extracted from a study on transit deserts in the city of Charlotte and the Charlotte/Mecklenburg Quality of Life Explorer, highlights the strong correlation between low-income communities and transit gaps [41, 42, 43, 44].

In contrast, East Charlotte fares somewhat better, with residents enjoying relatively better financial situations and access to urban amenities. However, challenges with urban public transportation and bus access persist in this area. North Charlotte, a more suburban region, mirrors West Charlotte's issues, particularly regarding transit dispari- ties caused by long distances between locations. This car-oriented area experiences more noticeable transportation challenges compared to other parts of Charlotte.

#### *3.2.2. Technology Integration: The CATS Mobile Application*

**App Features and Functionality:** The Charlotte Area Transit System (CATS) offers a mobile application known as "CATS-Pass," available for both iPhone and Android platforms. This app is designed to enhance the transit experience by providing a range of features that cater to the needs of modern commuters. The CATS-Pass app allows users to:

The app allows users to plan trips by entering start and destination points, providing route options with transfer details and travel times. It offers real-time bus tracking to minimize waiting, along with up-to-date schedule information and alerts for service changes. Mobile ticketing is available, enabling riders to purchase and store tickets digitally, supporting various payment methods. Additionally, the app provides notifications for service disruptions, helping users stay informed about any potential travel issues [45].

While the CATS-Pass app provides many essential features, users have reported some issues with the accuracy of real-time tracking and schedule information. These inaccuracies can lead to confusion and delays for riders, highlighting the need for continuous improvements in the app's data accuracy and reliability [46].

**Comparison with Other Transit Apps:** When comparing the CATS-Pass app with other popular transit apps like Transit, Moovit, and Google Maps, several differences become apparent:

- **Transit Apps:** Known for its user-friendly interface and robust real-time tracking, the Transit app not only provides accurate real-time data but also suggests alternative routes in case of delays. It also features integration with ride- sharing services and bike-sharing options, offering a comprehensive mobility solution [47].

- **Moovit:** Moovit excels in providing multimodal trip planning, incorporating public transit, ride-sharing, cycling, and walking options. It also offers live navigation with step-by-step directions, including alerts when it's time to get off the bus. The app's global reach and extensive data coverage make it a popular choice for international travelers [48].

- **Google Maps:** Google Maps offers a seamless transit experience with integrated real-time data, route planning, and walking directions. Its strength lies in its integration with other Google services and the ability to switch between driving, transit, cycling, and walking modes effortlessly. Google Maps also includes predictive travel times based on historical data and real-time traffic conditions [49].

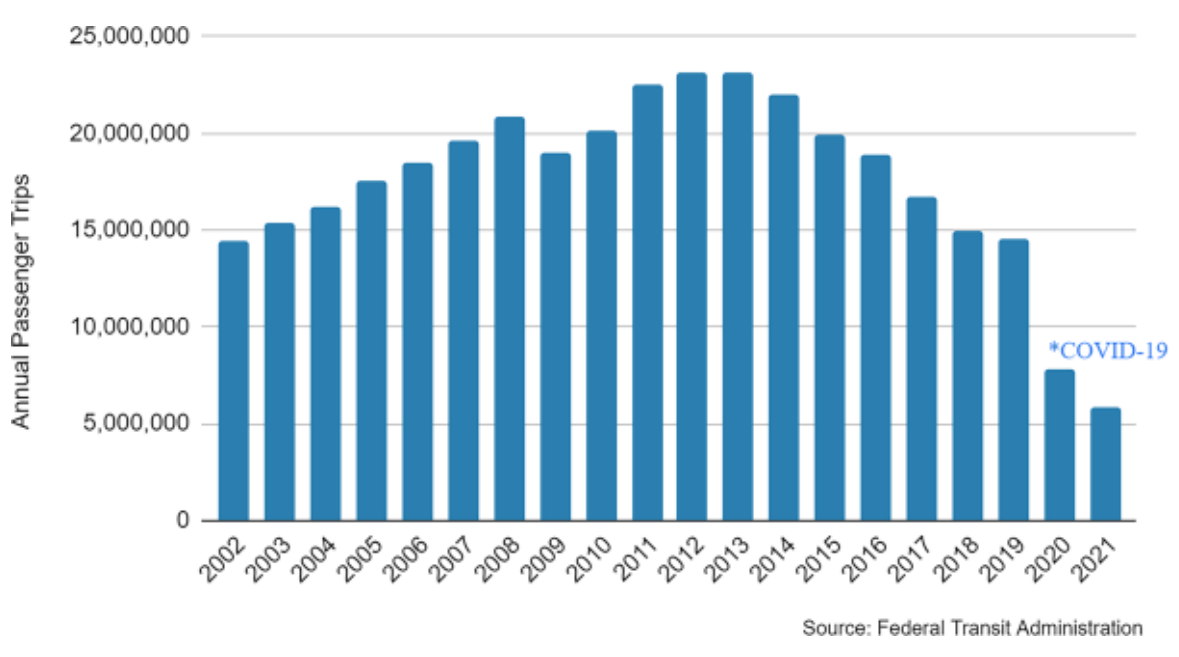

**Figure 3:** CATS Annual local bus ridership trend

Compared to these apps, CATS-Pass is more focused on the specific needs of Charlotte's transit system but lacks some of the advanced features and broader integration found in apps like Transit and Moovit. Enhancements in real-time tracking accuracy, multimodal options, and user interface improvements could make CATS-Pass more competitive and user-friendly.

### *3.2.3. Challenges and Future Directions for CATS Declining Ridership and Strategic Responses*

Charlotte is a predominantly car-dependent city, with 76.6% of its workforce driving alone to work, and only 3.4% using public transit. Contributing factors include Charlotte's sprawling growth pattern, the lack of a connected multimodal network, and the inefficiencies within the existing public transit system. Despite the variety of bus routes and services offered by CATS, public transportation remains an unpopular choice for many residents, especially outside of the low-income population who depend on it as their primary mode of transportation [7].

One of the critical issues facing Charlotte's bus transit system is declining ridership (Figure 3). Since its peak in 2013, bus ridership has steadily fallen, reaching its lowest level in 2021, with a 75% drop since 2014. This decline is the largest among the nation's 50 largest transit systems [28, 50]. However, CATS is actively seeking solutions to reverse this trend and improve the vibrancy and usefulness of the bus transit system.

### *3.2.4. On-Demand Services and Future Plans*

To attract riders back to the system, especially post-COVID, CATS is considering integrating more on-demand options and replacing low-ridership fixed-schedule routes with services that passengers could summon on-demand [51]. This initiative includes integrating on-demand services into low-income neighborhoods to enhance access to transit centers and popular bus routes, potentially through partnerships with ride-share companies, and bike and scooter services. One proposed strategy is to replace low-ridership circulator shuttles, like the North Meck Village Rider, with more flexible, on- demand services. As of now, the Village Rider has seen a significant decrease in ridership, with 32,393 riders so far this year, down almost 10% from last year and about half of its pre-COVID numbers. CATS aims to create a high-frequency network of buses that operate every 15 minutes or less, improving the frequency of core routes while managing low-ridership routes more effectively [51].

### *3.2.5. Long-Term Transportation Planning*

Charlotte is actively developing new strategies and proposals to address the current challenges in its public transportation system, particularly in bus transit. The *Charlotte Moves Task Force Report* emphasizes the need for an expanded network of high- frequency bus routes, with service intervals of 15 minutes or better on 22 more routes [52]. Riders have also expressed a strong desire for real-time bus arrival information at stops, as highlighted in the Bus Priority Study Report [53].

The *Connect Beyond Regional Mobility Plan* outlines the use of emerging mobility technologies and services to support the goal of creating connected and on-demand transit systems. One innovative goal includes the introduction of "Ride-hailing" services, where drivers and passengers connect via digital applications for pre-arranged and on-demand transportation services [54].

Additionally, the *Charlotte Future 2040 Comprehensive Plan*, a key document guiding the city's future, empha- sizes strengthening technology and partnerships to better manage congestion through advanced planning, intelligent transportation systems, demand management, and shared public/private funding strategies [55]. These objectives align closely with the goals of this study, which seeks to explore and propose innovative transit solutions for Charlotte.

## 4. Research Methodology

This section outlines the research methodology employed in this study, focusing on the selected CATS bus lines in the city of Charlotte.

The study investigates passengers' experiences and perceptions through surveys, aiming to gather data on public transit use, concerns, and the potential acceptance of an advanced mobile applications designed to enhance transit efficiency.

### 4.1. Study Area and Selected Bus Lines

The study focuses on the following CATS bus lines in Charlotte, with their routing maps shown in Figure 4:

- **Airport Sprinter Bus Line:** Connects City Center to the Charlotte Douglas International Airport. This bus route provides access to job opportunities, health services, and grocery stores near the uptown area for West Charlotte community residents. However, it primarily serves essential workers who work at the airport but live in low-income housing developments, forcing them to commute downtown first.

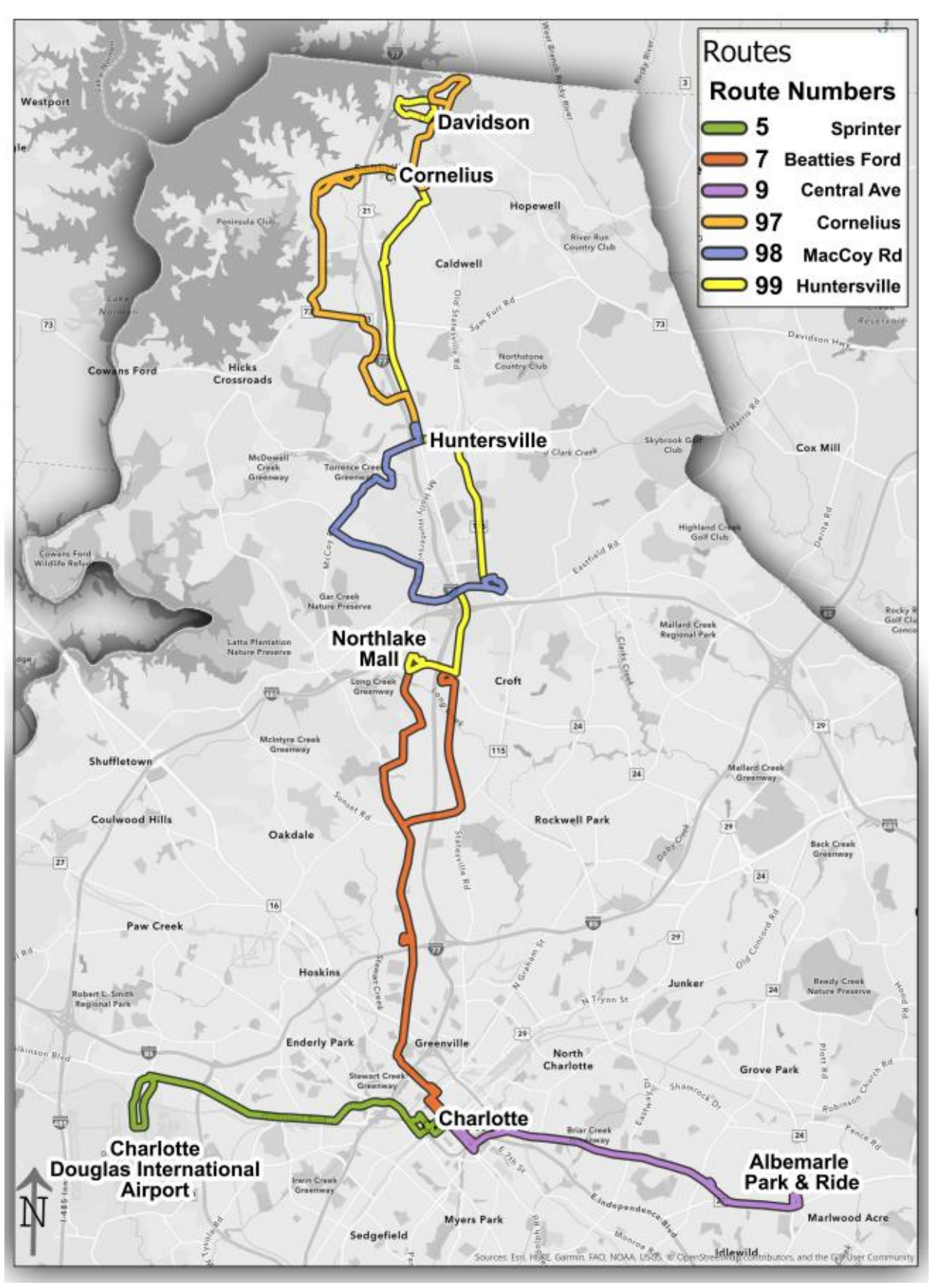

**Figure 4:** Case study CATS bus routes in the city of Charlotte

- **Bus Line 7:** Operates from the city center to Beatties Ford Road, a major thoroughfare in North Charlotte.
- **Bus Line 9:** Moves from the city transportation center to Central Avenue, the main street of East Charlotte, offering access to numerous urban amenities.
- **North Meck Village Rider (Lines 97-99):** Serves suburban areas in North Charlotte. known as North Meck Village Rider, which travel daily between Cornelius, Huntersville, and Davidson. The Village Rider routes connect to several CATS fixed route services at the Davidson-Gateway Park and Ride and the Northlake Mall Park and Ride. The North Meck Village Rider can serve destinations up to 3/4 mile off the main route, and passengers can use this service by contacting customer service agents one day in advance to schedule their trip.

## 4.2. Survey Methodology

The study employed a pencil-paper survey method, where the research team interacted with passengers at bus stops across the selected routes. Participants were provided with surveys to complete, focusing on their experiences and opinions regarding the current transit system and the potential for new technology. In addition to the survey, behavioral observation was employed as a qualitative research method. This approach helped to validate the survey findings by observing passenger behavior, such as their interaction with existing mobile apps and their general usage patterns of the bus system.

## 4.3. Survey Timing and Sample Size

The surveys were conducted from the spring of 2022 through the end of the summer, specifically from April to August 2022. This period was chosen to capture a range of passenger experiences during different seasons. However, the timing of the survey coincided with ongoing COVID-19 concerns, which had a significant impact on bus ridership across Charlotte. As a result, finding participants for the survey study from among bus riders proved to be particularly challenging.

Due to the pandemic, many regular bus users had either reduced their use of public transportation or avoided it altogether, leading to lower ridership levels. This situation made it difficult to reach a robust sample size, as fewer passengers were available and willing to participate in the study. Despite these challenges, the research team managed to survey a total of 75 participants across all bus lines (30 participants from Sprinter Line, 15 participants from Line 7, 15 participants from Line 9, 15 participants from Lines 97-99).

The survey was carried out during both peak and off-peak hours to ensure a diverse range of responses, though the reduced ridership likely influenced the demographics and opinions of those who did participate. The research acknowledges that the unique circumstances of the COVID-19 pandemic may have introduced biases in the data, particularly in terms of health-related concerns and the general willingness of passengers to use public transit. These factors are considered in the analysis of the survey results.

## 4.4. Survey Content and Questions

The survey was structured into four main sections, each designed to address specific research objectives. Tables 1 and 2 present the survey questions in the order they were asked in our questionnaire.

1- **Demographic Information:** This section gathered basic demographic data, including age range, gender, area of residence, and income level. The purpose was to understand the demographic profile of bus passengers on the selected routes.

2- **Current Transit Concerns:** Participants were asked about their experiences and concerns related to the current bus system. This section aimed to identify the shortcomings and problems within the existing transit infrastructure, such as waiting times, access to bus stops, and overall satisfaction with the service.

3- **Mobile Application Usage:** This section focused on participants' use of existing transit apps, particularly the CATS mobile application. Questions addressed the

**Table 1**
Survey Questions and Response Options

| Section | Question | Response Options |
|---|---|---|
| **Demographic Information** | | |
| **Demographics** | Please select your age range: | a) 18-24, b) 25-34, c) 35-44, d) 45-54, e) 55-64, f) 65-74, g) 75 or older |
| | Please select your gender: | a) Male, b) Female, c) Transgender, d) Other |
| | Please select your race/ethnicity: | a) African American/Black, b) White, c) Spanish, Hispanic or Latino, d) Asian/Pacific Islander, e) American Indian/Alaskan Native, f) Multi-racial, g) Other |
| | Please select the area that you live in or staying at: | a) North Charlotte, b) South Charlotte, c) East Charlotte, d) West Charlotte, e) City Center, f) Outside of local area |
| | Please select your income level: | a) Less than $12,500, b) $12,500 to $45,000, c) $45,000 to $80,000, d) Above $80,000 |
| **Current Transit Concerns** | | |
| **Bus Usage** | How often do you take the bus as transportation? | a) Daily, b) About once a week, c) About once a month, d) Rarely, e) Never |
| | What is the purpose of the trip you are currently taking? (Bus lines 7, 9, 97-99) | a) Work, b) Education, c) Medical, d) Leisure/Social/Recreation, e) Shopping/Errands/Groceries, f) Church, g) Other |
| | What is the purpose of taking the Sprinter bus? | a) To access a grocery store for shopping, b) To access the CLT Airport - As a worker, c) To access the CLT Airport - As a passenger, d) To access your workplace (Other than airport), e) To access other amenities such as healthcare, community centers, etc. |
| **Wait and Commute Times** | During the day and night, on average, how long do you usually have to wait for the bus to arrive at the bus stop? | a) 5-10 minutes, b) 10-20 minutes, c) 20-30 minutes, d) More than 30 minutes |
| | What is your commute time to get to the bus stop? | a) 5-10 minutes, b) 10-20 minutes, c) 20-30 minutes, d) More than 30 minutes |
| | How do you get to the bus station? | a) Walk, b) Another bus, c) Biking, d) Drop off |
| | What is the average time actually you spent on the bus? | a) 5-10 minutes, b) 10-20 minutes, c) 20-30 minutes, d) More than 30 minutes |
| **COVID-19 Concerns** | Considering the COVID-19 outbreak, how concerned are you about the health aspect of using public transportation or bus transit? | a) Not important at all, b) Usually concerned, c) Very concerned |
| | If you are concerned, please identify the reasons behind it: | a) Using overcrowded bus, b) Lack of wearing face masks in the bus, c) Not maintaining social distancing in the bus |
| **Reservation Services (Bus Lines 97-99)** | Have you ever made a scheduled reservation to use these buses to desired destinations? | a) Yes, b) No |
| | If your answer to the previous question is “No”, what is your major reason(s)? | a) Not being aware of this reservation service for planning a trip, b) Time limit for making a reservation, 24-hour prior to the trip, c) Difficulty of the reservation process, d) Lack of trust in the reservation service providers, e) The current system doesn’t meet your needs well, f) Others - please explain |
| | If you have an experience of making a scheduled reservation, has the bus arrived at your pick-up point according to your requested schedule? | a) Yes, b) No |
| | How satisfied are you with the reservation services of this line? | a) Very satisfied, b) Somewhat satisfied, c) Somewhat dissatisfied, d) Dissatisfied |

**Table 2**
Survey Questions and Response Options

| Section | Question | Response Options |
|---|---|---|
| **Reservation Services (Bus Lines 97-99)** | If you have used this reservation service and are not satisfied, please identify your major reason(s) for dissatisfaction: | a) Mismatch of the pick-up or drop-off points with the reservation and schedule requested, b) The bus did not arrive at the requested pick-up or drop-off points at the scheduled time, c) Problems in the process of calling customer service or filling out the form to reserve, d) Difficulty of making a reservation, one day before a trip (24-hour prior to a trip), e) Others - please explain |
| | If the scheduled reservation of the bus for a desired trip can be in real-time and in the next 10-30 minutes (more responsive to your needs), would you be willing to use it for your desired routes and destinations? | a) Yes, b) No |
| | What is your desired timeline to schedule a trip by the bus reservation service? | a) 10 minutes, b) 30 minutes, c) 1 hour, d) More than 1 hour to Less than 24 hours |
| **Ticket/Pass Purchasing** | Where did you purchase the ticket/pass you're using for your trip with this bus? | a) CTC (Transit Center Outlet), b) Ticket Machine, c) Mobile App, d) Employer, e) Online, f) Others |
| | **Mobile Application Usage** | |
| **Mobile Application Usage** | Do you own a smartphone and use it for planning your trips? | a) Yes, b) No |
| | If your answer to the previous question is "yes," which application do you use? | a) Google Maps, b) CATS mobile application, c) Other (Please specify its name) |
| | If you are using CATS mobile apps, how satisfied are you with this application? | a) Very satisfied, b) Somewhat satisfied, c) Somewhat dissatisfied, d) Dissatisfied |
| | If you are not satisfied, please identify your major reason(s) for dissatisfaction: | a) The bus schedules in the application are inaccurate, b) The application is not user-friendly, c) The application is slow, d) Others (Please explain) |
| | **Hypothetical Technology Acceptance** | |
| **Hypothetical Technology** | If such an application can reduce your wait time by 70% and reduce your overall trip time by 50%, would you be willing to use it? | a) Yes, b) No |
| | If your answer to the previous question is "No", what is your major concern? | a) Privacy, b) Lack of trust in urban transport authorities and applications, c) Lack of trust in the mobile application managers, d) The current system doesn't meet your needs well |
| | If such an application actually exists, how willing would you be to tap your travel information and use that application to get to your destination faster? | a) Very willing, b) Somewhat willing, c) Somewhat unwilling, d) Unwilling |

frequency of app usage, satisfaction levels, and specific issues or limitations faced by users.

4- **Hypothetical Technology Acceptance:** The final section introduced a new, hypothetical mobile application designed to optimize bus transit. Participants were asked about their willingness to use such technology, potential concerns (e.g., privacy, trust in the application), and their openness to sharing travel information to enhance route planning and scheduling.

### 4.5. Data Analysis

The data collected from the surveys and behavioral observations were primarily analyzed using descriptive methods. Quantitative data was summarized to provide insights into passenger demographics, usage patterns, and satisfaction levels. Additionally, descriptive analysis of open-ended responses was conducted to capture passengers' concerns and expectations regarding the current and future transit systems.

## 5. Results

In this section, the results of the passenger surveys conducted across the selected bus routes in Charlotte are presented. The data is segmented by urban (Sprinter, Line 7, Line 9) and suburban (Lines 97-99) lines to provide clearer insights. Also, the results are summarized in tables and figures for clarity. Table 3, summarizes the most frequent survey results for Demographics, Bus Usage, Wait and Commute Times, COVID-19 Concerns questions of this study.

### 5.1. Part One: Demographics

The demographic section of the survey aimed to capture the age, gender, residential area, and income level of the participants. The results are segmented into urban and suburban categories.

#### *5.1.1. Urban Lines (Sprinter, Line 7, Line 9)*

The urban lines showed a distribution of age groups with a significant representation from the working-age population (25-64 years). The majority of participants were male, reflecting a trend across all urban lines.

- Age Distribution: On the Sprinter Bus Line, the most common age group was 55-64 years (23%), followed by the 25-34 age group (20%). On Bus Line 7, 40% of respondents were aged 35-44, while Bus Line 9 had 53% of participants aged 45-64.

- Gender Distribution: The majority of respondents across all urban lines were male, with the highest male representation on Line 9 (86.7%). Sprinter had 73.3% male participants, while Line 7 had 80%.

- Residential Area: Participants predominantly lived in underserved communities of North and West Charlotte for the Sprinter and Bus Line 7. Line 9's participants were mostly from East Charlotte.

- Income Level: A large proportion of participants reported an income level between $12,500 to $45,000, indicating that many users of these bus lines come from underserved communities.

#### *5.1.2. Suburban Lines (Lines 97-99)*

For the suburban lines, the age distribution was slightly different, with a noticeable presence of older participants (65-74 years). Again, males were the predominant gender among participants.

- Age Distribution: Participants were either working-age (35-44 years) or nearing retirement (65-74 years).

- Gender Distribution: Two-thirds of the respondents were male (66.7%).

- Residential Area: Participants mainly resided in North Charlotte, where the suburban lines operate. This area had a higher percentage of African American/Black participants (80%).

- Income Level: Like the urban lines, most participants reported low-income levels, with a substantial portion earning between $12,500 to $45,000.

### 5.2. Part Two: Quality of Travel

This section focuses on participants' experiences with bus travel, including wait times, commute times, concerns related to COVID-19, and the purpose of their trips.

#### *5.2.1. Urban Lines (Sprinter, Line 7, Line 9)*

Participants on the urban lines reported moderate to long wait times, particularly during peak hours. The average time spent commuting to the bus stop ranged from 5 to 20 minutes, with most participants walking to the bus stop.

- Wait Times: On the Sprinter Bus Line, 43% of participants waited between 10-20 minutes for the bus. Bus Line 7 also had 40% of participants reporting the same wait time. Line 9 participants similarly indicated a wait time of 10-20 minutes during the day and night.

- Commute Times: Participants generally spent 10-30 minutes on the bus, with Sprinter and Line 9 participants noting this range as the average time on the bus. Line 7 had a similar range, with most participants spending 10-20 minutes on the bus.

- COVID-19 Concerns: A significant number of participants expressed concerns about COVID-19, with overcrowding being the primary issue. About 80% of Sprinter participants were concerned about health aspects, while 48% on Line 7 were worried about overcrowding.

- Purpose of the Trips:

  Sprinter Bus Line: The Sprinter line primarily serves participants commuting for work (67%), particularly those employed at or near the airport. A significant portion (29%) also used the bus for grocery shopping and accessing other essential services.

  Bus Line 7: Participants on Bus Line 7 primarily used the bus for work (60%), with others using it to access amenities such as healthcare facilities, community centers, and grocery stores.

  Bus Line 9: Bus Line 9 had a strong focus on work-related trips (67%), with a smaller percentage using it for other purposes, including accessing social services and educational institutions.

#### *5.2.2. Suburban Lines (Lines 97-99)*

The suburban lines had longer wait times due to their lower ridership and the greater distances between stops.

- Wait Times: Most participants on the suburban lines reported waiting 5-20 minutes for the bus. However, some reported wait times as long as an hour.

- Commute Times: Commutes were longer compared to urban lines, with many participants spending more than 20 minutes on the bus.

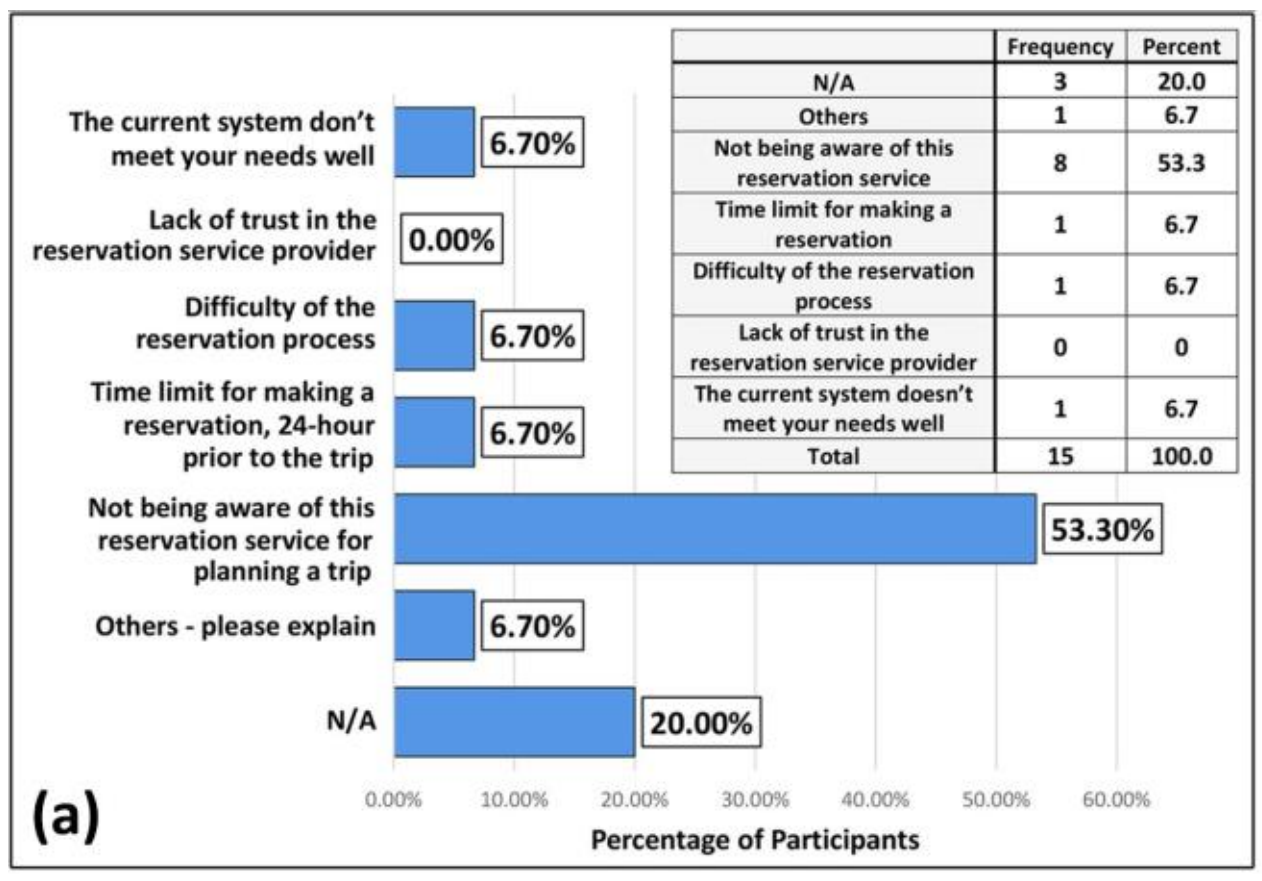

| | Frequency | Percent |
|---|---|---|
| N/A | 3 | 20.0 |
| Others | 1 | 6.7 |
| Not being aware of this reservation service | 8 | 53.3 |
| Time limit for making a reservation | 1 | 6.7 |
| Difficulty of the reservation process | 1 | 6.7 |
| Lack of trust in the reservation service provider | 0 | 0 |
| The current system doesn't meet your needs well | 1 | 6.7 |
| Total | 15 | 100.0 |

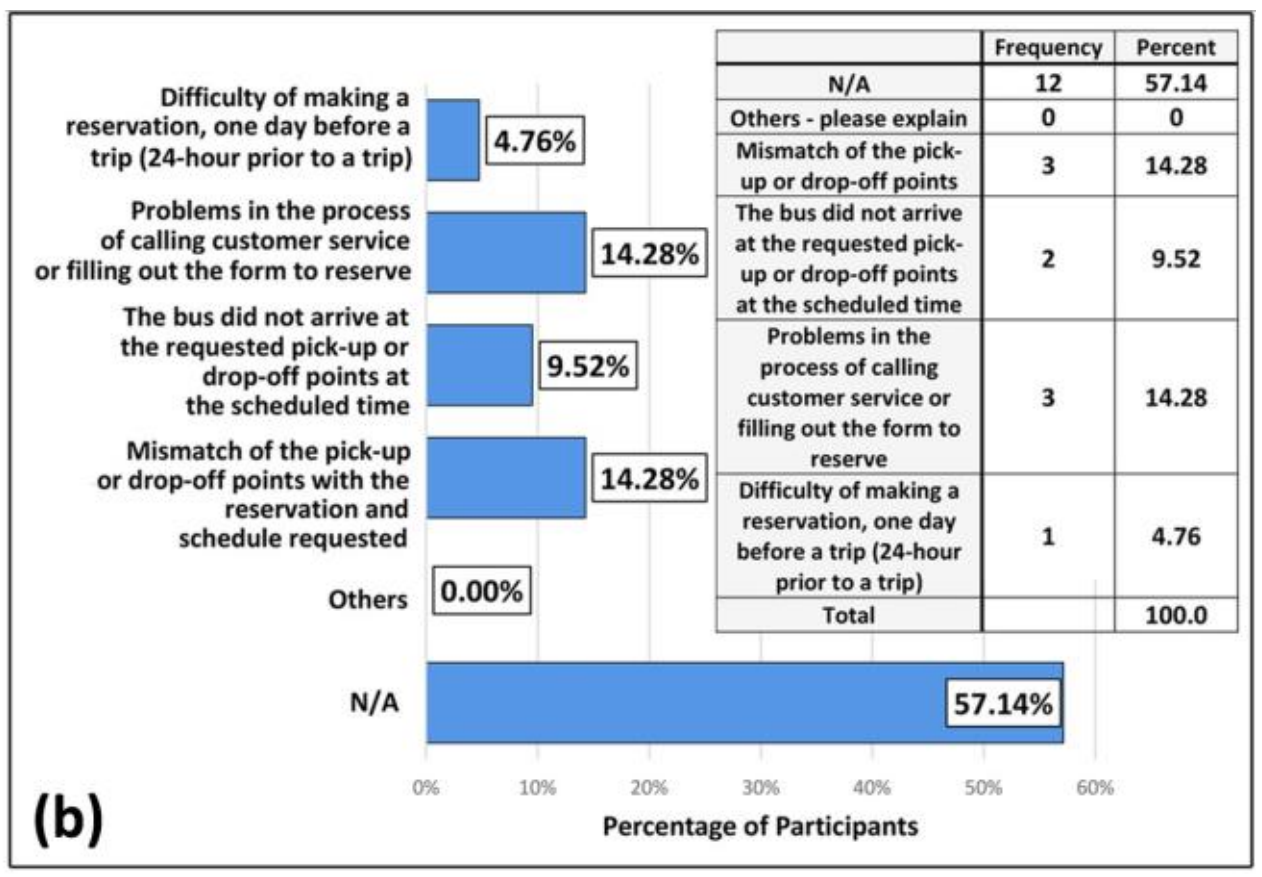

| | Frequency | Percent |
|---|---|---|
| N/A | 12 | 57.14 |
| Others - please explain | 0 | 0 |
| Mismatch of the pick-up or drop-off points | 3 | 14.28 |
| The bus did not arrive at the requested pick-up or drop-off points at the scheduled time | 2 | 9.52 |
| Problems in the process of calling customer service or filling out the form to reserve | 3 | 14.28 |
| Difficulty of making a reservation, one day before a trip (24-hour prior to a trip) | 1 | 4.76 |
| Total | | 100.0 |

**Figure 5:** (a) The major reasons for not having an experience of using the reservation system - (Bus Lines 97-99), (b) If participants have used this reservation service and they are not satisfied with it, what is their major reason(s) for dissatisfaction - (Bus Lines 97-99)

- COVID-19 Concerns: Participants were generally less concerned about COVID-19, possibly due to lower bus occupancy. Only a small percentage reported significant concerns.
- Purpose of the Trips: The suburban routes showed a more diverse range of purposes. While 47% used the bus for work, a substantial percentage also relied on the service for accessing healthcare, community services, and shopping (33%). These routes are crucial for connecting residents to essential services outside of their immediate residential areas.

**Reservation System Experience (Bus Lines 97-99):**
The reservation system is currently only available for Bus Lines 97-99. According to the survey results, only 3 out of 15 participants had used the reservation system.

Reasons for Not Using the Reservation System: The majority of those who had not used the system cited a lack of awareness as the main reason (53.3%). Other reasons included the time limit for making a reservation (6.7%) and difficulty in using the system (6.7%) (Figure 5 (a)).

Dissatisfaction with the Reservation System: Among those who had used the reservation system, the primary reasons for dissatisfaction were that the bus did not arrive at the requested pick-up or drop-off points at the scheduled time (14.3%), and problems with calling customer service or filling out the form to reserve (14.3%). The difficulty of making a reservation one day before departure was also cited as a reason for dissatisfaction (4.8%) (Figure 5 (b)).

## 5.3. Part Three: CATS Mobile Application Usage

This section explores participants' usage of the current CATS mobile application and their satisfaction levels.

### *5.3.1. Urban Lines (Sprinter, Line 7, Line 9)*

A substantial portion of participants owned smartphones and used the CATS mobile application. However, dissatisfaction was prevalent due to inaccurate bus schedules.

- Smartphone Ownership: A majority of participants on the urban lines owned smartphones, with over half using the CATS app. Sprinter had 67% smartphone ownership.
- Satisfaction Levels: Dissatisfaction stemmed mainly from inaccurate schedules, which hindered effective trip planning. About 33% of Sprinter participants and a similar percentage of Line 7 participants expressed dissatisfaction with the CATS app due to schedule inaccuracies.

The measurement of satisfaction and dissatisfaction was conducted through a survey that included specific questions related to the accuracy of bus schedules, the usability of the CATS mobile application, and the overall quality of the transit service. Satisfaction levels were quantified based on these responses, with a focus on identifying key areas of concern for passengers.

The current survey results indicate that dissatisfaction is a main issue, particularly due to the inaccuracies in bus schedules. This dissatisfaction was expressed by 27-33% of participants, depending on the specific bus line (Figure 6).

### *5.3.2. Suburban Lines (Lines 97-99)*

Smartphone ownership was lower among suburban line users, and the CATS app was less frequently used, with participants favoring Google Maps.

- Smartphone Ownership: 60% of participants owned smartphones, and fewer participants used the CATS app compared to urban lines.

**Table 3**
Summary of the Most Frequent Survey Results for All Bus Lines - (Demographics, Bus Usage, Wait and Commute Times, COVID-19 Concerns)

| Bus Line | Description | Data/Observation |
|---|---|---|
| **Sprinter Bus Line** | Gender | 73.3% Male, 26.7% Female |
| | Age range | 13.3% (18-24), 20% (25-34), 16.7% (35-44), 13.3% (45-54), 23.3% (55-64), 13.3% (65-74), - (75 or above) |
| | Participants living area in Charlotte | Predominantly West and North Charlotte |
| | Participants income level | $12,550 to $45,000 |
| | Purpose of the trip | 60% use bus daily, 29% for grocery shopping |
| | How participants get to the bus station | 63% used another bus to reach the station |
| | Waiting time for the bus | 43% wait 10-20 minutes |
| | Time spent on the bus | 67% spend 10-30 minutes |
| | COVID-19 concerns | 80% concerned, 50% concerned about overcrowding |
| **Bus Line 7** | Gender | 80.0% Male, 20.0% Female |
| | Age range | 6.7% (18-24), 26.7% (25-34), 40.0% (35-44), 13.3% (45-54), 6.7% (55-64), 6.7% (65-74), - (75 or above) |
| | Participants living area in Charlotte | North Charlotte |
| | Participants income level | $12,550 to $45,000 |
| | Purpose of the trip | 60% for work, 20% for accessing other amenities |
| | How participants get to the bus station | 50% walk, 50% use another bus |
| | Waiting time for the bus | 40% wait 10-20 minutes |
| | Time spent on the bus | 67% spend 10-20 minutes |
| | COVID-19 concerns | 47% not concerned, 53% concerned about crowding |
| **Bus Line 9** | Gender | 86.7% Male, 13.3% Female |
| | Age range | 13.3% (18-24), 20.0% (25-34), 13.3% (35-44), 26.7% (45-54), 26.7% (55-64), - (65-74), - (75 or above) |
| | Participants living area in Charlotte | East Charlotte |
| | Participants income level | $12,550 to $45,000 |
| | Purpose of the trip | 67% for work |
| | How participants get to the bus station | 67% walk |
| | Waiting time for the bus | 67% wait 10-20 minutes |
| | Time spent on the bus | 80% spend 10-30 minutes |
| | COVID-19 concerns | 47% not concerned, 53% concerned about crowding |
| **Bus Lines 97-99** | Gender | 66.7% Male, 33.3% Female |
| | Age range | - (18-24), 20.0% (25-34), 26.7% (35-44), 6.7% (45-54), 20.0% (55-64), 26.7% (65-74), - (75 or above) |
| | Participants living area in Charlotte | North Charlotte |
| | Participants income level | $12,550 to $45,000 |
| | Purpose of the trip | 47% for work |
| | How participants get to the bus station | 50% walk, 30% use another bus, 20% bike |
| | Waiting time for the bus | 80% wait 5-20 minutes |
| | Time spent on the bus | 67% spend 10-30 minutes |
| | COVID-19 concerns | 40% concerned, 60% not concerned |

- Satisfaction Levels: Despite lower usage, those who did use the app were generally dissatisfied, citing inaccurate schedules as a key issue. Only 30% used the CATS app for navigation, preferring Google Maps instead.

## 5.4. Part Four: Willingness to Adopt New Technology

This final section assesses participants' willingness to adopt a new, smart on-demand transit application. Before this section of the survey, we first explained a hypothetical

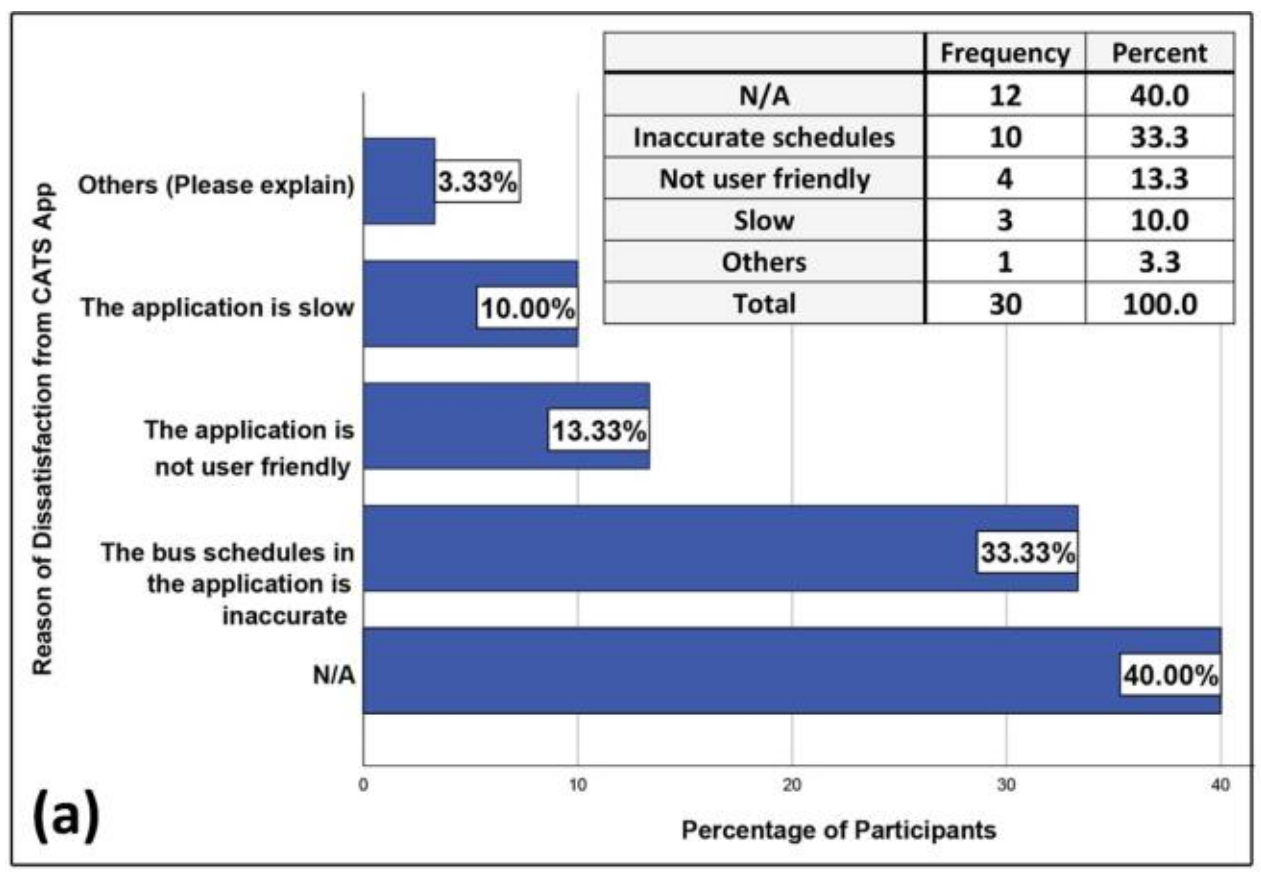

| | Frequency | Percent |
|---|---|---|
| N/A | 12 | 40.0 |
| Inaccurate schedules | 10 | 33.3 |
| Not user friendly | 4 | 13.3 |
| Slow | 3 | 10.0 |
| Others | 1 | 3.3 |
| Total | 30 | 100.0 |

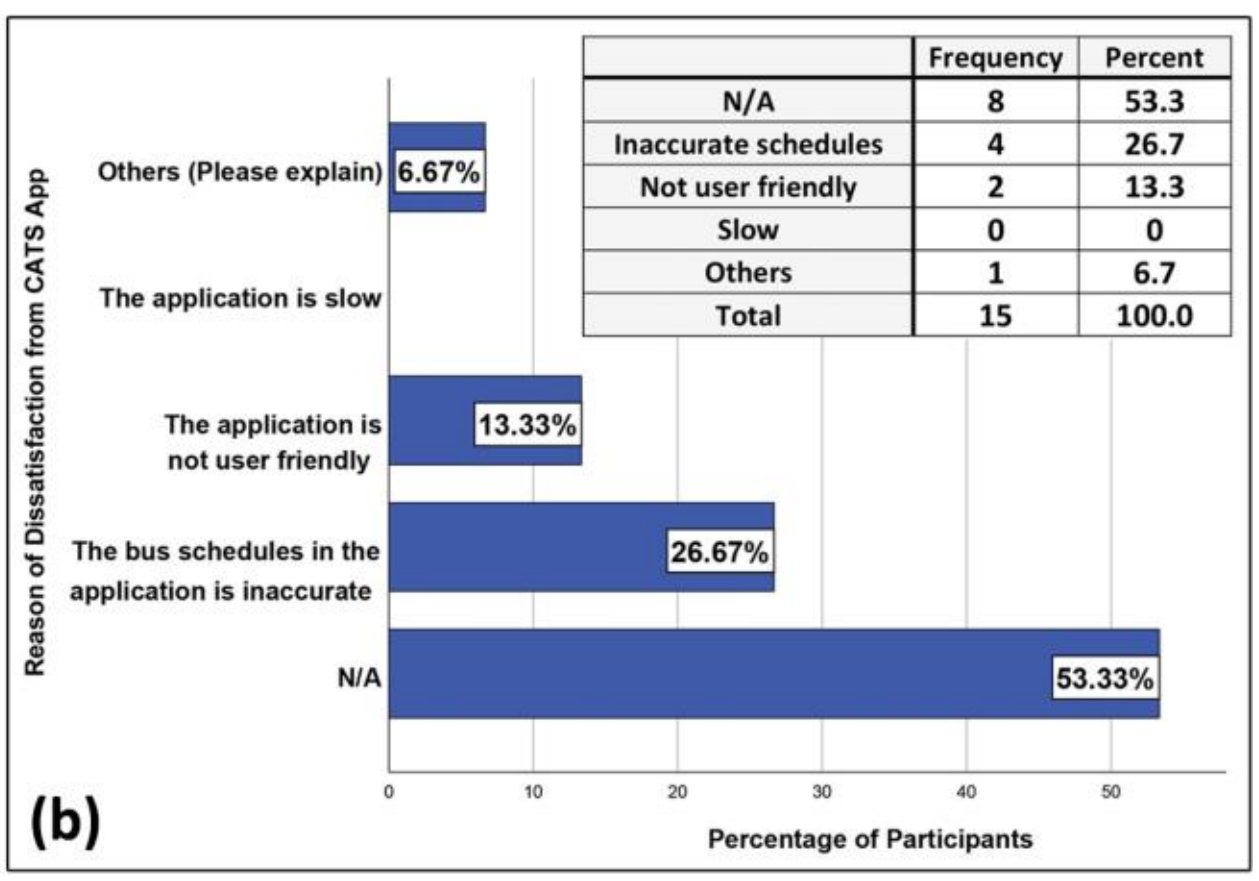

| | Frequency | Percent |
|---|---|---|
| N/A | 8 | 53.3 |
| Inaccurate schedules | 4 | 26.7 |
| Not user friendly | 2 | 13.3 |
| Slow | 0 | 0 |
| Others | 1 | 6.7 |
| Total | 15 | 100.0 |

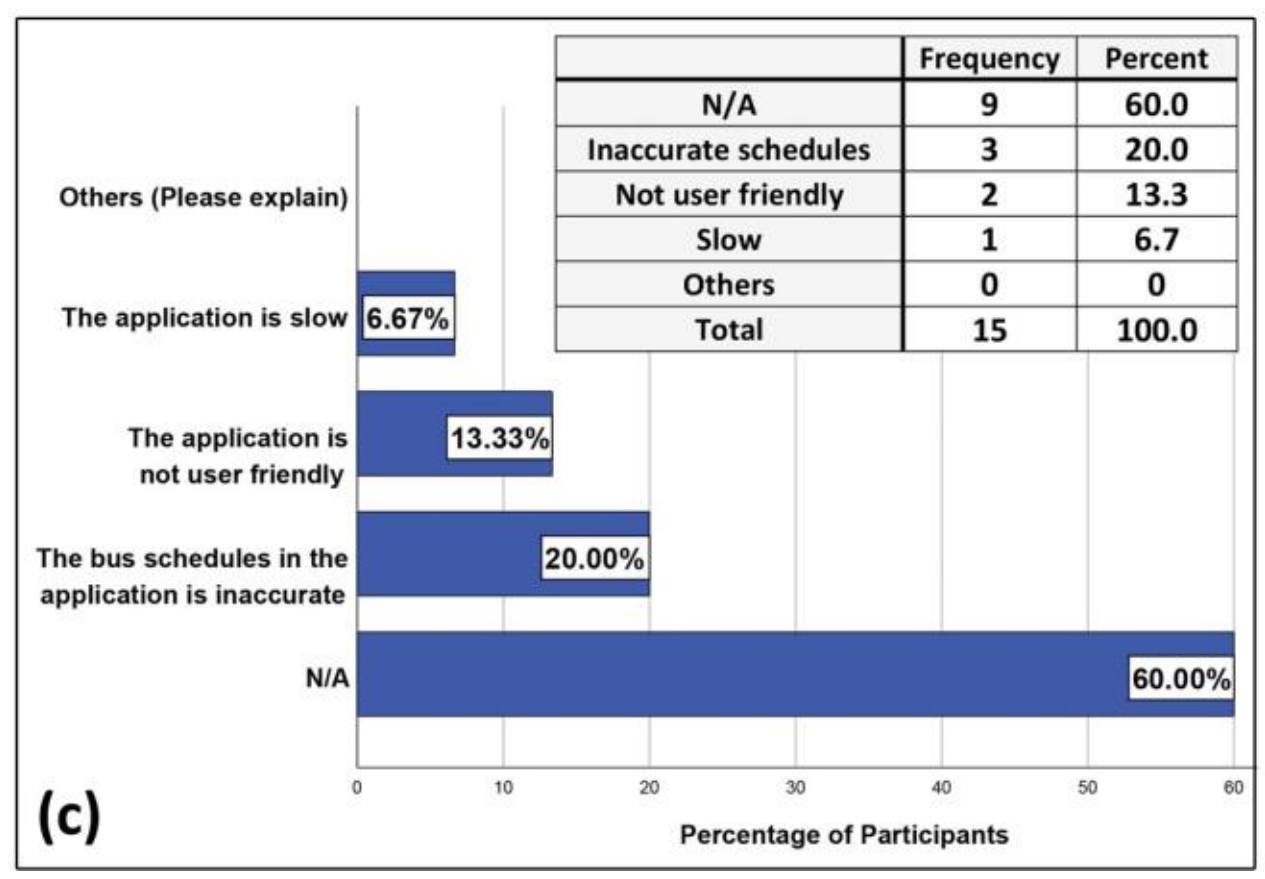

| | Frequency | Percent |
|---|---|---|
| N/A | 9 | 60.0 |
| Inaccurate schedules | 3 | 20.0 |
| Not user friendly | 2 | 13.3 |
| Slow | 1 | 6.7 |
| Others | 0 | 0 |
| Total | 15 | 100.0 |

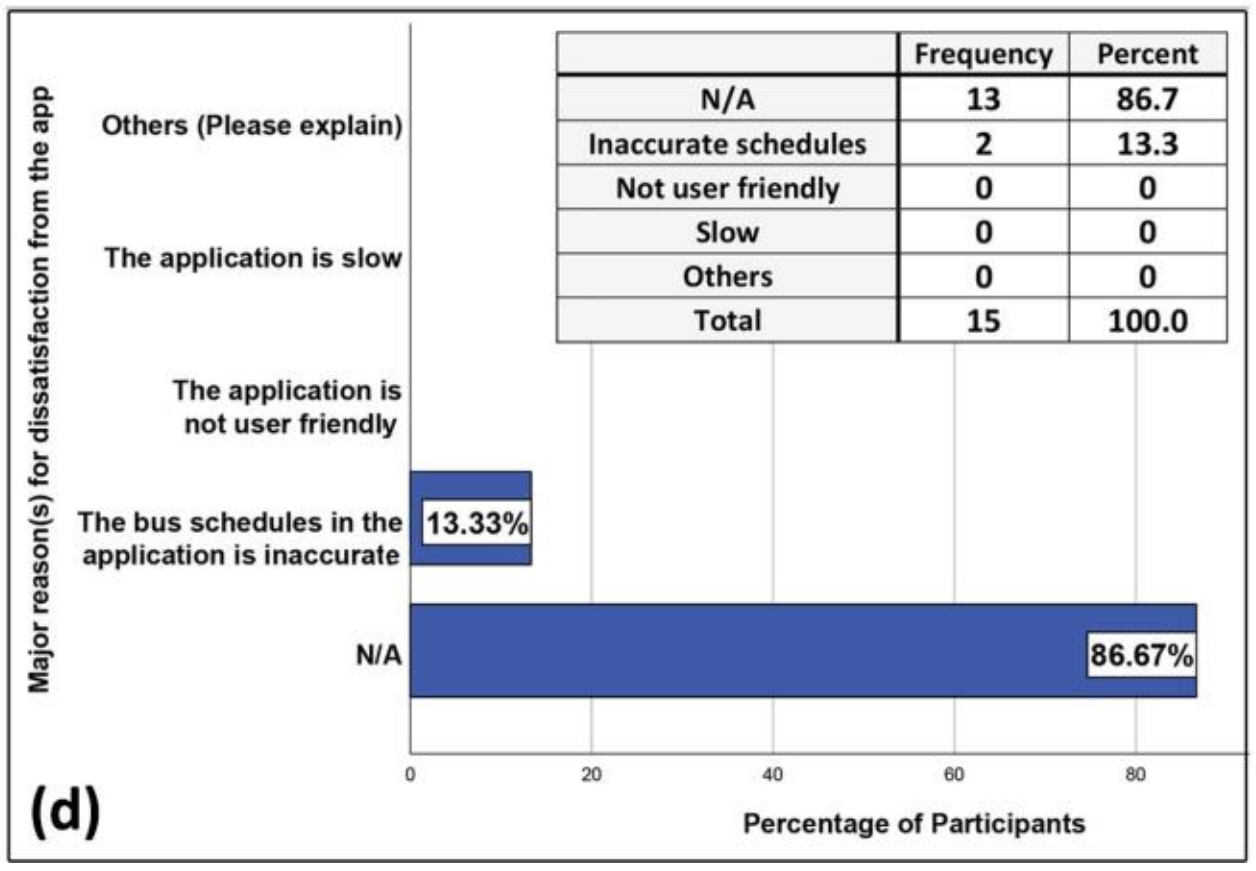

| | Frequency | Percent |
|---|---|---|
| N/A | 13 | 86.7 |
| Inaccurate schedules | 2 | 13.3 |
| Not user friendly | 0 | 0 |
| Slow | 0 | 0 |
| Others | 0 | 0 |
| Total | 15 | 100.0 |

**Figure 6:** Participants' major reason(s) for dissatisfaction with the CATS mobile application, if they are not satisfied: (a) Sprinter Bus Line, (b) Bus Line 7, (c) Bus Line 9, (d) Bus Lines 97-99

technology to participants with a statement as follows:

*The new technology be a novel mobile application akin to the Uber application, with the difference that here, the application will only serve bus transit. In this application, riders will be able to enter their desired destination. According to riders' origin, by intelligently examining the bus stations and routes leading to their destination, the application will tell riders, in a customized way, which bus to take to reach that destination. Arguably, our novel technology will make the use of city buses more desirable and efficient for people. It helps riders reach their destination in the shortest time possible, efficiently, and easily.*

#### 5.4.1. Urban Lines (Sprinter, Line 7, Line 9)

The majority of participants expressed willingness to adopt the new application, provided it could significantly reduce wait and commute times.

- Willingness to Adopt: 90% of Sprinter participants and 100% of Line 9 participants were willing to use the new application. Line 7 also showed high willingness at 86%.

- Concerns: Privacy was the most common concern, particularly among those hesitant to adopt new technology.

#### 5.4.2. Suburban Lines (Lines 97-99)

Participants on suburban lines also showed interest in the new application, particularly if it could offer real-time reservation services.

- Willingness to Adopt: 93% of participants were willing to use the new application, highlighting the potential for improving transit services.

- Concerns: Privacy remained a key concern, similar to the urban lines.

The survey results highlight that while the current bus system remains crucial for low-income communities in Charlotte, there are significant opportunities for enhancement, particularly in service reliability, application accuracy, and responsiveness to real-time demands. Although participants show a strong willingness to embrace new, smarter transit solutions for different bus lines shown in Figure 7, this adoption hinges on addressing their concerns, especially regarding privacy and trust in the technology.

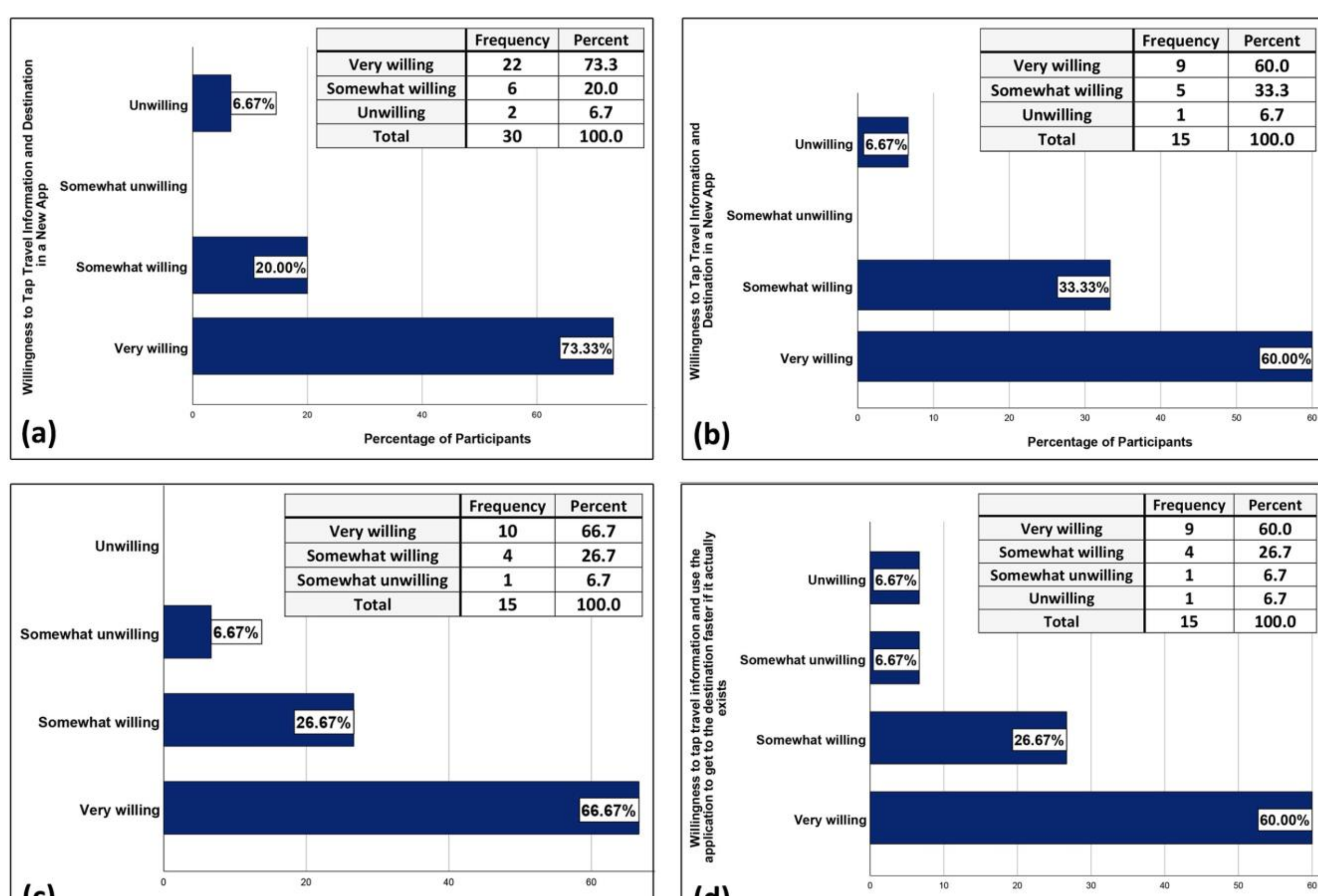

**Figure 7:** If such an application actually exists, how willing would participants be to tap their travel information and use that application to get to their destination faster: (a) Sprinter Bus Line, (b) Bus Line 7, (c) Bus Line 9, (d) Bus Lines 97-99

Ensuring these issues are adequately managed is vital for the successful implementation of future public transportation improvements in Charlotte.

## 6. Discussion

This study offers a comprehensive analysis of Charlotte's public bus transit system, providing insights into current service levels and exploring the potential for innovative, data-driven improvements. The results of passenger surveys across various bus lines highlight key areas for enhancement and underscore the importance of addressing the needs of underserved communities. The introduction of a proposed smart, on-demand transit application represents a promising direction for the future of public transportation in Charlotte and similar cities.

### 6.1. Benchmarking Current Service and Potential Improvements

This study provides a benchmark for evaluating the current performance of Charlotte's bus system, highlighting areas for improvement such as schedule inaccuracies, long wait times, and dissatisfaction with the CATS mobile application. The proposed smart on-demand technology could address these issues, offering a significant upgrade to the existing system. Although some findings, like overall satisfaction with the current CATS mobile application, did not reach statistical significance, they still provide valuable insights. The widespread dissatisfaction with schedule accuracy underscores a critical area for improvement that could substantially impact overall service quality.

Moreover, the analysis of purpose-driven trips—such as commuting for work, grocery shopping, and accessing essential services—highlights the diverse needs of bus riders and also the transit and service gaps existing in some areas. For instance, while the Sprinter bus line is specifically designed to facilitate travel between Uptown Charlotte and the CLT airport, some participants from West Charlotte use this route to access grocery stores and other essential amenities. The smart on-demand application could be tailored to optimize these specific types of trips, ensuring that the most common routes are prioritized for efficiency. This reinforces the need for continuous innovation in public transit to meet the evolving needs of the population, particularly those in underserved communities.

## 6.2. Sociodemographic and Behavioral Influences on Technology Adoption

Survey results indicate that willingness to adopt the new smart transit application is influenced by various sociodemographic factors and travel habits. Younger passengers, particularly those aged 18-34, show a higher propensity to use mobile applications and express interest in the new technology. Frequent bus users who experience longer wait and commute times are also more likely to adopt the application, as it promises significant improvements in their transit experience.

However, the results also reveal significant gender and age disparities that could influence the adoption of new technology. For instance, the predominance of male riders, particularly on Bus Line 9, suggests that the design and communication strategies for the new technology should consider the specific needs and preferences of this demographic. Additionally, the varied age distributions across bus lines indicate that while younger passengers are more likely to adopt new technology, older passengers, who may be less tech-savvy, might need additional support and reassurance, particularly regarding privacy concerns and data security.

An essential consideration is the percentage of passengers who do not own a smartphone. This segment of the population, which may include a significant portion of low-income or older passengers, presents a challenge for the adoption of app-based DRPT systems. For these passengers, the benefits of the system may seem limited. However, the proposed smart system includes video surveillance of buses and stops connected to a centralized system, ensuring that all passengers can benefit from on-time bus transit, even if they do not have direct access to the mobile application. Addressing these concerns through clear communication and user-friendly design will be crucial to the system's success.

## 6.3. Linking Survey Findings to Future Developments

The final question in the survey, which assessed participants' willingness to use the proposed smart transit application, effectively links current service levels with potential future developments. The strong support for the new technology, despite some reservations, indicates a demand for more efficient, responsive, and user-friendly transit solutions. This aligns with the broader goals of the study, which aim to propose innovative transit solutions that better serve Charlotte's residents, particularly those in underserved communities.

The study also highlights the importance of educating the public about new transit technologies, especially in areas where there is a lack of awareness or understanding of existing systems, such as the reservation system currently available for Bus Lines 97-99. Providing clear information and support will be essential to ensuring widespread adoption and satisfaction with the new system.

## 6.4. Future of Demand-Responsive Public Transit (DPRT)

The introduction of smart, on-demand technology represents a significant shift in how public transit systems could operate in the future. This technology, as proposed in the survey statement briefly and shown in Figure 8, would function similarly to ride-hailing services like Uber but would be tailored specifically for bus transit. By using real-time data and smart algorithms, the system would optimize bus routes, reduce wait times, and enhance the overall efficiency of public transit. This vision aligns with the growing trend toward demand-responsive public transit (DPRT), where transit services are dynamically adjusted based on real-time passenger demand.

The proposed technology is not just a theoretical concept but a practical solution that could be implemented within the existing infrastructure of Charlotte's bus system. By leveraging current resources such as cameras, sensors, and GPS technology, the system could be integrated with minimal cost and disruption, offering a model that could be replicated in other cities. This approach would not only improve service for current transit-dependent populations but also attract new users by offering a more reliable and efficient alternative to private vehicle use.

## 6.5. Transferability of Results

The findings from Charlotte's bus system may have broader implications for similar urban and suburban areas in the United States and beyond. The issues identified—such as long wait times, inefficient routes, and dissatisfaction with existing mobile applications—are not unique to Charlotte. These challenges are common in many cities, especially those with underserved communities that rely heavily on public transportation. Thus, the results of this study could serve as a valuable reference for other U.S. locations with comparable demographic and transit conditions, offering a roadmap for improving bus transit systems elsewhere.

Internationally, cities with comparable socioeconomic and transit dynamics, especially in developing countries or areas with emerging transit systems, can also benefit from the insights gained from this study. The proposed smart on-demand technology offers a scalable solution that can be adapted to different contexts, potentially transforming public transit on a global scale. By addressing common issues such as schedule inefficiencies and long wait times, this technology could significantly improve the quality of public transit in diverse settings.

## 6.6. Policy Implications and Future Directions

As cities like Charlotte look toward the future of public transit, the findings of this study provide a valuable foundation for policy development and strategic planning. The successful implementation of smart, on-demand technology in Charlotte could serve as a model for other cities, demonstrating how data-driven solutions can enhance public transit efficiency, reduce operational costs, and improve service quality. Additionally, by focusing on the needs of transit-dependent populations, this technology has the potential to contribute to greater social equity in urban transportation.

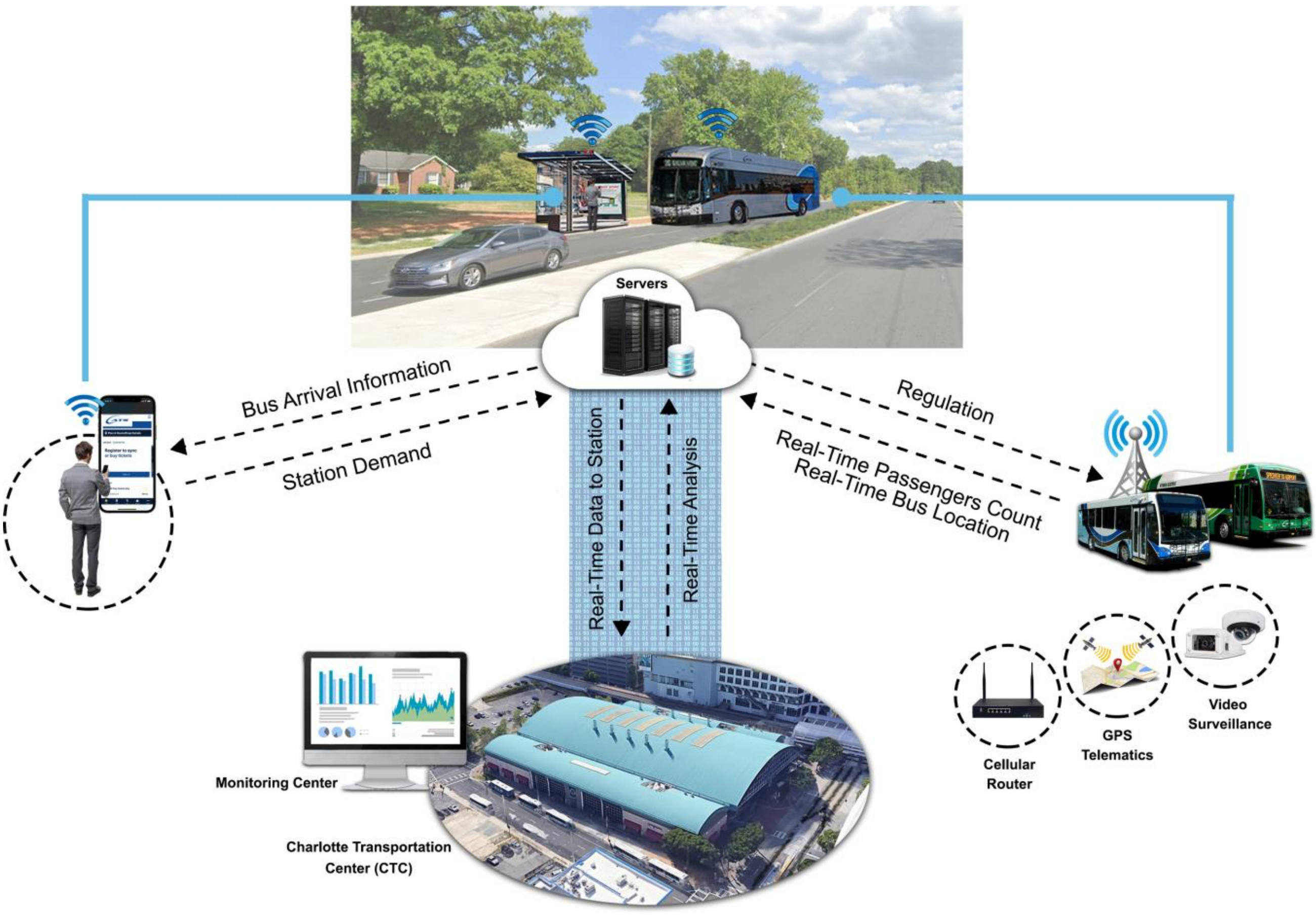

**Figure 8:** Future proposed technology for the smart on-demand bus system in the city of Charlotte

The study's findings suggest that public transit systems must continue to evolve to meet the needs of a diverse and changing population. This evolution requires not only technological advancements but also thoughtful policy decisions that prioritize accessibility, equity, and sustainability. Ultimately, the goal is to create a more efficient, reliable, and equitable transit system that better serves all residents, positioning public transit as a viable and attractive alternative to private vehicles.

## 7. Conclusion

This study focused on examining the current habits and the acceptability of Demand-Responsive Public Transit (DRPT) within low- income areas of Charlotte, with a particular emphasis on underserved communities. The primary objective was to identify key challenges in the existing public transit system and assess the potential for integrating smart, on-demand technologies to improve service quality and accessibility.

The findings reveal that Charlotte's bus system faces significant issues, particularly with long wait times, inefficient routing, and dissatisfaction with current transit applications. These problems are particularly pronounced in low-income areas, where transit- dependent populations experience greater challenges. However, while there is a general interest in adopting new technologies, the majority of survey participants expressed concerns related to privacy, trust, and the effectiveness of such systems, particularly among older and low-income passengers. This suggests that while there is potential for DRPT systems to improve public transit, significant barriers to adoption remain, particularly in underserved communities.

The study also highlights that the proposed smart, on-demand transit technology could serve as a valuable tool for addressing these challenges, but its success will depend heavily on addressing the concerns of the target population. The technology's ability to optimize bus routes, reduce wait times, and enhance overall efficiency aligns with broader trends in DRPT, but its implementation in Charlotte must be carefully tailored to meet the specific needs of its diverse and often vulnerable populations. While the results from Charlotte provide valuable insights, their applicability to other cities, particularly those with similar transit dynamics, should be approached cautiously. The challenges and opportunities identified in Charlotte may well be relevant to other car-centric cities in the southern United States, such as Austin, TX; Atlanta, GA; and Nashville, TN. However, the transferability of these findings should be tested further, considering the unique socio-economic and geographic characteristics of each city.

In conclusion, this study underscores the need for a careful, context-sensitive approach to the implementation of smart DRPT systems in Charlotte and similar cities. While there is a clear demand for improved transit services, particularly in low-income areas, the success of these innovations will depend on addressing the specific concerns and needs of the target population. This research contributes to a better understanding of the existing gaps and potential future desires in the field of smart, connected, and on-demand bus transit systems, laying the groundwork for future studies and implementations aimed at creating more equitable and efficient public transit solutions.

## Acknowledgements

This research has been funded by the Gambrell Faculty Fellowship Program in 2021-2022 and is supported by Charlotte Area Transit System (CATS). The authors would like to thank the Charlotte Area Transit System for its support and assistance throughout the research process and all the other people who contributed to the survey process.

## References

[1] T. Ghosh, T. Kanitkar, R. Srikanth, Assessing equity in public transportation in an indian city, Journal of Transport Geography 102 (2022) 103299. doi:10.1016/j.jtrangeo.2022.103299. URL https://www.sciencedirect.com/science/article/pii/S2213624X22001948

[2] S. Shaheen, A. Cohen, Mobility on demand (mod) and mobility as a service (maas): early understanding of shared mobility impacts and public transit partnerships, in: E. Deakin (Ed.), Transportation, Land Use, and Environmental Planning, Elsevier, 2019, Ch. 3, pp. 37–56. doi:10.1016/B978-0-12-815018-4.00003-6. URL https://www.sciencedirect.com/science/article/pii/B9780128150184000036

[3] F. Faghihinejad, M. M. Zoghifard, A. M. Amiri, S. Monajem, Evaluating social and spatial equity in public transport: A case study, Urban Policy and Research 41 (1) (2022) 1–18. doi:10.1080/19427867.2022.2158541. URL https://www.tandfonline.com/doi/full/10.1080/19427867.2022.2158541

[4] AllTransit, Alltransit performance score, https://alltransit.cnt.org/ (2023).

[5] R. Brough, M. Freedman, D. Phillips, Understanding socioeconomic disparities in travel behavior during the covid-19 pandemic, Journal of Regional Science 61 (4) (2021) 753–774.

[6] K. L. Covington, Overcoming spatial mismatch: The opportunities and limits of transit mode in addressing the black-white unemployment gap, City & Community 17 (1) (2018) 211–235.

[7] S. Hosseini, M. Azarbayjani, H. Tabkhi, Community-driven approach for smart on-demand public transit in charlotte in underserved communities: Pilot study for user acceptance and early data collection, in: Proceedings of the ARCC 2023 International Conference - The Research-Design Interface, Dallas, TX, 2023, pp. 25–32, available: https://www.researchgate.net/publication/378964143.

[8] J. J. C. Aman, J. Smith-Colin, Transit deserts: Equity analysis of public transit accessibility, Transportation Research Part A: Policy and Practice 151 (2021) 236–252. doi:10.1016/j.tra.2021.07.018. URL https://www.sciencedirect.com/science/article/pii/S0966692320309467

[9] A. Unknown, Bridging digital divides: a literature review and research agenda for information systems research, Information Systems Frontiers 24 (3) (2021) 599–615. doi:10.1007/s10796-020-10096-3. URL https://link.springer.com/article/10.1007/s10796-020-10096-3

[10] E. V. Research, Ai-controlled bus transit services in japan, https://www.electricvehiclesresearch.com/articles/16773/ai-controlled-us-transit-services-in-japan (2023).

[11] E. I. Diab, M. G. Badami, A. M. El-Geneidy, Bus transit service reliability and improvement strategies: Integrating the perspectives of passengers and transit agencies in north america, Transport Reviews 35 (3) (2015) 292–328.

[12] TSG, Transport, data analytics and ai: Why tfl's latest initiative is good news, https://www.tsg.com/insights/transport-data-analytics-and-ai-why-tfls-latest-initiative-is-good-news/ (2022).

[13] J. R. J. Öberg, A. G. M. Glaumann, G. L. Berntsson, Smart public transport, http://ci.nii.ac.jp/naid/40006345637/ (2017).

[14] C. T. Systems, Nextbus, https://www.cubic.com/sites/default/files/2020-02/Cubic-CTS-brochure-nextbus-V3.pdf (2020).

[15] ResearchGate, A multi-agent reinforcement learning approach for bus holding control strategies, https://www.researchgate.net/publication/303575786_A_Multi-Agent_Reinforcement_Learning_approach_for_bus_holding_control_strategies (2016).

[16] S. M. H. Moosavi, A. Ismail, C. W. Yuen, Using simulation model as a tool for analyzing bus service reliability and implementing improvement strategies, PloS ONE 15 (5) (2020) e0232799.

[17] W. Chen, C. Yang, F. Feng, Z. Chen, An improved model for headway-based bus service unreliability prevention with vehicle load capacity constraint at bus stops, Discrete Dynamics in Nature and Society 2012 (2012) Article ID 975620.

[18] B. Yu, J.-b. Yao, Z.-Z. Yang, An improved headway-based holding strategy for bus transit, Transportation Planning and Technology 33 (3) (2010) 329–341.

[19] F. Sun, A. Dubey, J. White, A. Gokhale, Transit-hub: A smart public transportation decision support system with multi-timescale analytical services, Cluster Computing 22 (1) (2019) 2239–2254.

[20] K. Gkiotsalitis, R. Kumar, Bus operations scheduling subject to resource constraints using evolutionary optimization, in: Informatics, Vol. 5, 2018, p. 9.

[21] K. Koh, C. Ng, D. Pan, K. S. Mak, Dynamic bus routing: A study on the viability of on-demand high-capacity ridesharing as an alternative to fixed-route buses in singapore, in: 2018 21st International Conference on Intelligent Transportation Systems (ITSC), 2018, pp. 34–40.

[22] S. Nannapaneni, A. Dubey, Towards demand-oriented flexible rerouting of public transit under uncertainty, in: Proceedings of the Fourth Workshop on International Science of Smart City Operations and Platforms Engineering, 2019, pp. 35–40.

[23] N. A. of City Transportation Officials, Transit route types, https://nacto.org/publication/transit-street-design-guide/introduction/service-context/transit-route-types/ (2023).

[24] I. Transport, Automatic passenger counting systems for public transport, https://www.intelligenttransport.com/transport-articles/3116/automatic-passenger-counting-systems-for-public-transport/ (2022).

[25] D. Morales, J. C. Muñoz, P. Gazmuri, A stochastic model for bus injection in an unscheduled public transport service, Transportation Research Part C: Emerging Technologies 113 (2020) 277–292.

[26] R. Sensing, Automatic passenger counting, https://www.retailsensing.com/automated-passenger-counting.html/ (2023).

[27] Markets, Markets, Automated passenger counting system market by technology application - covid-19 impact analysis, https://www.marketsandmarkets.com/Market-Reports/automatic-passenger-counting-information-system-market-17856390.html (2022).

[28] U. of North Carolina at Charlotte Urban Institute, Charlotte considers replacing some low-ridership buses with on-demand service, https://ui.charlotte.edu/story/charlotte-considers-replacing-some-low-ridership-buses-demand-service (2023).

[29] K. Harmony, V. Gayah, User-centered design in transit apps: Enhancing usability and real-time information, in: Proceedings of the Transportation Research Board 96th Annual Meeting, Washington, D.C., 2017, pp. 1–15.

[30] C. Mulley, J. D. Nelson, S. Wright, D. A. Hensher, Digital platforms and public transportation: The transformation of services through technology, Transportation Research Part A: Policy and Practice 104 (2017) 111–122.

[31] J. C. Romero, M. Iacono, D. M. Levinson, Factors influencing the adoption of mobile transit apps in smart cities, Journal of Urban Technology 29 (3) (2022) 35–52.

[32] A. Inemek, P. Matthyssens, The impact of buyer-supplier relationships on supplier innovativeness: An empirical study in cross-border supply networks, Industrial Marketing Management 42 (4) (2013) 580–594.

[33] V. Vasseur, R. Kemp, The adoption of pv in the netherlands: A statistical analysis of adoption factors, Renewable and Sustainable Energy Reviews 41 (2015) 483–494.

[34] X. Liu, Q. Wang, W. Wang, Evolutionary analysis for residential consumer participating in demand response considering irrational behavior, Energies 12 (19) (2019) 3727.

[35] F. Corsini, C. Certomà, M. Dyer, M. Frey, Eparticipatory energy: Research, imaginaries and practices on people's contribution to energy systems in the smart city, Technological Forecasting and Social Change 142 (2019) 322–332.

[36] D. Marikyan, S. Papagiannidis, E. Alamanos, A systematic review of the smart home literature: A user perspective, Technological Forecasting and Social Change 138 (2019) 139–154.

[37] Wikipedia, Charlotte area transit system, https://en.wikipedia.org/wiki/Charlotte_Area_Transit_System (2023).

[38] C. A. T. System, Transit vision 2030, https://charlottenc.gov/cats/newsroom/Documents/2030-Transit-Vision.pdf (2023).

[39] C. A. T. System, Charlotte riders guide, https://charlottenc.gov/cats/bus/Documents/Charlotte-Riders-Guide.pdf (2023).

[40] C. of Charlotte, Charlotte area transit system – fast facts, https://charlottenc.gov/cats/newsroom/Pages/fast-facts.aspx (2023).

[41] A. Tirachini, O. Cats, Covid-19 and public transportation: Current assessment, prospects, and research needs, Journal of Public Transportation 22 (1) (2020) 1–21.

[42] R. Chetty, N. Hendren, P. Kline, E. Saez, Where is the land of opportunity? the geography of intergenerational mobility in the united states, The Quarterly Journal of Economics 129 (4) (2014) 1553–1623.

[43] J. Jiao, M. Dillivan, Transit deserts: The gap between demand and supply, Journal of Public Transportation 16 (3) (2013) 23–39.

[44] M. Q. o. L. E. Charlotte, Quality of life explorer, https://mcmap.org/qol/#24/ (2023).

[45] C. A. T. S. (CATS), Cats mobile apps, https://www.charlottenc.gov/CATS/Bus/Mobile-Apps (2023).

[46] C. A. T. System, Fare equity analysis for fy2017 proposed fare increase february 2016, https://charlottenc.gov/cats/about/Documents/CATSFY17FareIncreaseEquityAnalysisFinal.pdf (2016).

[47] I. Transit App, Transit app: Live bus and train schedules, real-time transit information, https://transitapp.com/ (2023).

[48] Moovit, Moovit: Public transit and mobility app, https://moovitapp.com/ (2023).

[49] G. LLC, Google maps: Navigation and transit, https://maps.google.com/ (2023).

[50] WUNC, Getting off the bus: Cats has plans to bring riders back after massive drop. will it work?, https://www.wunc.org/2022-07-19/ (2022).

[51] WBTV, Cats to integrate on-demand services, expand to low-income neighborhoods, https://www.wbtv.com/2022/06/06/cats-integrate-uber-like-service-expand-low-income-neighborhoods/ (2022).

[52] C. Observer, Commuter rail essential from charlotte to lake norman, former mayor says, https://account.charlotteobserver.com/paywall/subscriber-only?resume=238858703&intcid=ab_archive (2022).

[53] C. A. T. System, Envision my ride - bus priority study - final report, https://charlottenc.gov/cats/transit-planning/envisionmyride/Documents/View-the-Bus-Priority-Study-Final-Report-here.pdf (2022).

[54] C. R. Council, Connect beyond, regional mobility plan, https://www.connect-beyond.com/docs/CONNECT_Beyond_Final_Report_Final.pdf (2022).

[55] C. of Charlotte, Charlotte future 2040 comprehensive plan, https://cltfuture2040.com/ (2023).